\documentclass[useAMS,usenatbib]{mn2e}

\usepackage{graphicx}
\usepackage{txfonts}
\usepackage{epsfig}
\usepackage{tabularx}
\usepackage{subfig}

\title[Formation of blue hook stars in GCs - II]
{A possible formation channel for blue hook stars in globular cluster - II.
Effects of metallicity, mass ratio, tidal enhancement efficiency  and helium abundance }
\author[Lei et al.]{Zhenxin Lei$^{1,2}$\thanks{lzx2008@ynao.ac.cn}, Gang Zhao$^{2}$\thanks{gzhao@nao.cas.cn},
Aihua Zeng$^{1}$, Lihua Shen$^{1}$, Zhongjian Lan$^{1}$
\and
Dengkai Jiang$^{3,4}$ and Zhanwen Han$^{3,4}$\thanks{zhanwenhan@ynao.ac.cn} \\
$^{1}$Department of Science, Shaoyang University, Shaoyang 422000, China\\
$^{2}$Key Laboratory of Optical Astronomy, National Astronomical Observatories, Chinese Academy of Sciences, Beijing 100012, China\\
$^{3}$Key Laboratory for the Structure and Evolution of Celestial Objects,
Chinese Academy of Sciences,Kunming 650011, China\\
$^{4}$Yunnan Observatory, Chinese Academy of Sciences, Kunming 650011, China}

\begin{document}

\date{Accepted ; Received ; in original form}

\pagerange{\pageref{firstpage}--\pageref{lastpage}} \pubyear{2015}

\maketitle

\label{firstpage}

\begin{abstract}
Employing tidally enhanced stellar wind, we studied in binaries the
effects of metallicity, mass ratio of primary to secondary, tidal enhancement
efficiency and helium abundance on the formation of blue hook (BHk) stars in
globular clusters (GCs). A total of 28 sets of binary models combined with different input parameters
are studied. For each set of binary model, we presented a range of initial orbital periods that is
needed to produce BHk stars in binaries. All the binary models could produce BHk stars
within different range of initial orbital periods. We also compared our
results with the observation in the $\it{T}\rm_{eff}$-$\rm{log}\it{g}$ diagram of
GC NGC 2808 and $\omega$ Cen. Most of the
BHk stars in these two GCs locate well in the region predicted by our
theoretical models, especially when C/N-enhanced model atmospheres are considered.
We found that mass ratio of primary to secondary and tidal enhancement
efficiency have little effects  on the formation of BHk stars in binaries, while
metallicity and helium abundance would play important roles, especially for
helium abundance. Specifically, with helium abundance increasing in binary models, the
space range of initial orbital periods needed to produce BHk stars becomes obviously wider,
regardless of other input parameters adopted. Our results were discussed with
recent observations and other theoretical models.
\end{abstract}


\section{Introduction}

Horizontal branch (HB) stars in globular clusters (GCs) are
low mass stars that are burning helium in their cores. These
stars are considered to be the progeny of red giant branch (RGB) stars (Hoyle \& Schwarzschild 1955).
To locate on different positions of HB, RGB stars need to lose
different envelope masses before or during helium core flash is
taking place (Catelan 2009). Some stars could settle on the red HB (RHB) positions in the
colour-magnitude diagram (CMD) of GCs after losing a few envelope masses,
while other stars may occupy the blue HB (BHB) or
extreme HB (EHB) positions due to which they lose most of or nearly the whole
envelope masses on the RGB stage. However, the physical mechanism of
mass loss for RGB stars in GCs is still unclear (Willson 2000; Dupree et al. 2009).

In the late 1990s, a special kind of hot EHB stars
are found in some massive GCs (e.g., NGC 2808, $\omega$ Cen; Whitney et al. 1998; D'Cruz et al. 2000),
which are the so called blue hook (BHk) stars.
These stars present very high temperatures
(e.g., $T\rm_{eff}$ $>$ 32000 K; Moni Bidin et al. 2012) and very faint luminosity
when compared with normal EHB stars in GCs. Therefore, BHk stars
can not be predicted by canonical stellar evolution models, and
their formation mechanism is still unclear. So far, several
formation scenarios are proposed for BHk stars in GCs (see Heber 2016 for a recent review).
D'Antona et al. (2010) proposed that BHk star in $\omega$ Cen
could be the progeny of blue main-sequence (MS) stars
that belong to the second generations in this GC (also see Lee 2005). These
stars would undergo an extra mixing during the RGB stage, thus
present very high helium abundance in their surfaces
(e.g., up to $Y\approx$ 0.8). After helium ignites in
their cores, these stars locate on very blue and faint
HB position and become BHk stars in GCs. On the other hand,
Brown et al. (2001) suggested that BHk stars could be
produced through late hot flash process on the
white dwarf (WD) cooling curve (also see
Castellani \& Castellani 1993;
D'Cruz et al. 1996; Brown et al. 2010, 2012).
In this scenario, low mass stars
in GCs experience huge mass loss on the RGB and
undergo helium core flash, instead of at the RGB tip,
on the way towards to
WD stage (early hot flash; Brown et al. 2001;
Cassisi et al. 2003; Miller Bertolami et al. 2008) or on the WD cooling
curve (late hot flash; Brown et al. 2001;
Cassisi et al. 2003; Miller Bertolami et al. 2008). During the late hot
flash, the internal convection can permeate the thin hydrogen-enriched
envelope and lead to helium and carbon enhancement
in the surface. Therefore, when settling on the
zero age horizontal branch (ZAHB), these stars that experience
late hot flash could
be hotter and fainter than normal EHB stars in GCs.

Recently, more and more pieces of evidence both from
photometry and spectroscopy support that
multiple populations could be a universal phenomenon
in most of the Galactic GCs (Piotto et al. 2007;
Gratton Carratta \& Bragaglia 2012; Gratton et al. 2013, 2014; Milone 2015)
and this phenomenon is considered to be closely
correlated with helium enrichment in GCs (D'Antona \& Caloi 2008;
Marino et al. 2014; Milone 2015; but see
Jiang et al. 2014 for an alternative solution).
If this is the case, stars in GCs would belong
to different populations that present different
helium abundances,
and BHk stars would be the progeny of helium-enriched
populations that belong to the second generation
stars in GCs (D'Antona et al. 2002; Brown et al. 2012).
Following this scenario,  Tailo et al. (2015) also
studied the effects of rapidly rotating second generation
stars on the formation of BHk stars in GCs.
They found that an increase of helium core mass up to
0.04 $M_{\odot}$ is required to solve the luminosity
range problem for BHk stars in $\omega$ Cen.

Lei et al. (2015, hereafter Paper I) followed
the late hot flash scenario and proposed that
tidally enhanced stellar wind in binary evolution (Tout \& Eggleton 1988)
is a possible formation channel for BHk stars in
GCs. This kind of wind could provide huge mass loss
on the RGB stage naturally, which is needed in
late hot flash scenario (Brown et al. 2001).
Their results indicated that binaries could
 produce BHk stars under tidally enhanced
stellar wind and it may play
important roles in the formation of BHk stars in some GCs.
However, Paper I did not study the effects of other input parameters
on their results, such as metallicity, mass ratio
of primary to secondary,
tidal enhancement efficiency, and helium abundance, etc.
These parameters would influence the evolution of  binary systems
 (e.g., mass loss of the primary, stellar mass, helium core mass, luminosity, etc), thus
may have influences on the formation of BHk stars in GCs.
As a further study for Paper I, to investigate
the role of binaries in the formation of BHk stars,
we studied in this paper the effects of
metallicity, mass ratio, tidal enhancement efficiency and
helium abundance on the binary evolution by
considering tidally enhanced stellar wind into binary evolution,
thus studied their effects on the formation of BHk stars in GCs.
The structure of this paper is organized as follows:
In Section 2, we describe the models and method used in this study;
results are given in Section 3; and finally, discussion and
conclusions are  given in Sections 4 and 5 respectively.

\section[]{Methodology}
As in Paper I, we use equation (1) to
describe the tidally enhanced stellar wind
in binary evolution, which was first suggested by
Tout \& Eggleton (1988).
\begin{equation}
\dot{M}=-\eta4\times10^{-13}(RL/M)\{1+B\rm_{w}\times \rm min[{\it(R/R\rm_{L})}^{6},
\rm 1/2^{6}]\},
\end{equation}
where $\eta$ is the Reimers mass-loss efficiency (Reimers 1975),
and $B\rm_{w}$ is the efficiency of tidal
enhancement for the stellar wind.  Here $R$, $L$, and $M$ are
the radius, luminosity and mass of the primary star in solar units.
Equation (1) was added into detailed stellar evolution code,
Modules for Experiments in Stellar Astrophysics ({\scriptsize MESA},
version 6208; Paxton et al. 2011, 2013, 2015) to study its effects on
the mass loss of primary during RGB evolution stage.

\begin{table*}
\small
 \begin{minipage}{80mm}
  \caption{Main input parameters used in the study.
  The masses of primary stars at ZAMS  in each set
  correspond to an age of about 12 Gyr at RGB tip.}
  \end{minipage}\\

\begin{tabular}{ccccccc}
  \end{tabular}\\
    \begin{tabularx}{9.4cm}{XcccccX}
    \hline\noalign{\smallskip}
 Model &$M_{\rm ZAMS}/M_{\odot}$ & $Y$ &$P_{1}/{\rm d}$  &$P_{2}/{\rm d}$ &$P_{3}/{\rm d}$ &$P_{4}/{\rm d}$\\
    \hline\noalign{\smallskip}
       I  &$Z$=0.003, &$q$=1.6, &$B\rm_{w}$=10000\\
          \hline\noalign{\smallskip}
  set 1 &  0.87 &  0.24 &  2850  & 2700 & 2160  & 2150  \\
  set 2 &  0.81 &  0.28 &  3100  & 2850 & 2230  & 2220  \\
  set 3 &  0.75 &  0.32 &  3500  & 3100 & 2330  & 2320  \\
  set 4 &  0.64 &  0.40 &  10000 & 4600 & 2680  & 2670  \\
    \hline\noalign{\smallskip}
         II  &$Z$=0.001, &$q$=1.6, &$B\rm_{w}$=10000\\
          \hline\noalign{\smallskip}
  set 5 &   0.83 &  0.24 &  2200  & 2000 & 1610  & 1600  \\
  set 6 &  0.77 &  0.28 &  2300  & 2150 & 1670  & 1660  \\
  set 7 &  0.72 &  0.32 &  2600  & 2260 & 1700  & 1690  \\
  set 8 &  0.62 &  0.40 &  10000  & 3100 & 1890  & 1880  \\
    \hline\noalign{\smallskip}
            III   &$Z$=0.0003, &$q$=1.6, &$B\rm_{w}$=10000\\
          \hline\noalign{\smallskip}
  set 9 &  0.81 &  0.24 &  1700  & 1560 & 1300  & 1290  \\
  set 10 &  0.758 &  0.28 &  1800  & 1650 & 1320  & 1310  \\
  set 11 &  0.70 &  0.32 &  2050  & 1800 & 1400  & 1390  \\
  set 12 &  0.61 &  0.40 &  10000  & 2350 & 1500  & 1490  \\
      \hline\noalign{\smallskip}
             IV    &$Z$=0.001, &$q$=1.2, &$B\rm_{w}$=10000\\
          \hline\noalign{\smallskip}
  set 13 &  0.83 &  0.24 &  2250  &2150 & 1730  & 1720  \\
  set 14 &  0.77 &  0.28 &  2600  & 2300 & 1800  & 1790  \\
  set 15 &  0.72 &  0.32 &  2650  & 2400 & 1820  & 1810  \\
  set 16 &  0.62 &  0.40 &  10000  & 3200 & 2020  & 2010  \\
    \hline\noalign{\smallskip}
             V     &$Z$=0.001, &$q$=2.4, &$B\rm_{w}$=10000\\
          \hline\noalign{\smallskip}
  set 17 &  0.83 &  0.24 &  1900  & 1750 & 1440  & 1430  \\
  set 18 &  0.77 &  0.28 &  2100  & 1900 & 1500  & 1490  \\
  set 19 &  0.72 &  0.32 &  2350  & 2050 & 1530  & 1520  \\
  set 20 &  0.62 &  0.40 &  10000  & 2750 & 1720  & 1710  \\
    \hline\noalign{\smallskip}
             VI    &$Z$=0.001, &$q$=1.6, &$B\rm_{w}$=5000\\
          \hline\noalign{\smallskip}
  set 21 &  0.83 &  0.24 &  1800  & 1680 & 1350  & 1340  \\
  set 22 &  0.77 &  0.28 &  1960  & 1780 & 1410  & 1400  \\
  set 23 &  0.72 &  0.32 &  2100  & 1900 & 1430  & 1420  \\
  set 24 &  0.62 &  0.40 &  10000  & 2700 & 1590  & 1580  \\
    \hline\noalign{\smallskip}
                VII     &$Z$=0.001, &$q$=1.6, &$B\rm_{w}$=1000\\
          \hline\noalign{\smallskip}
  set 25 &  0.83 &  0.24 &  1200  & 1130 & 910  & 900  \\
  set 26 &  0.77 &  0.28 &  1350  & 1200 & 940  & 930  \\
  set 27 &  0.72 &  0.32 &  1500  & 1260 & 960  & 950  \\
  set 28 &  0.62 &  0.40 &  10000  & 1750 & 1070  & 1060  \\
    \hline\noalign{\smallskip}
  \end{tabularx}
\end{table*}

We use the binary module in {\scriptsize MESA} to evolve the primary star in a binary system from
zero-age main sequence (ZAMS) to WD
cooling curve. Since the stellar wind on the RGB stage
could be tidally enhanced by the secondary stars,
some of the primary stars could lose
nearly the whole envelope mass and
evolve off RGB tip, then experience a late hot helium
flash on WD cooling curve (Castellani \& Castellani 1993;
D'Cruz et al. 1996; Brown et al. 2001). Due to the very thin hydrogen envelope,
internal convection mixing triggered by helium core flash is able to penetrate
into the surface and lead to helium and carbon enhancement.
When settling on HB, these stars could present higher
temperatures and faint luminosity than normal EHB star.

In this study, the default values of input physics in {\tiny MESA} are used
except for the opacity tables. {\scriptsize OPAL} type II tables
are used in our model calculations that would be more
suitable for helium burning stars (see Paper I).
 The Reimers mass-loss efficiency, $\eta$, is set to 0.45
(Renzini \& Fusi Pecci 1988; McDonald \& Zijlstra 2015).
All the input parameters used in this study are listed in Table 1.
Columns 1-3 gives the model number, stellar mass of the
primary star at ZAMS ($M_{\rm ZAMS}$) and initial helium abundance.
The value of $M_{\rm ZAMS}$ for each model adopted here corresponds
to an age of about 12 Gyr at RGB tip. Columns 4-7 present
some critical orbital periods for binary models, which will be introduced
in next paragraph.
Different from Paper I, we adopted three values of metallicity
(i.e., $Z$=0.003, 0.001 and 0.0003) in this study to
investigate its effects  on the final results.
For each metallicity, the mass ratio of primary to secondary
(i.e., $q$) and tidal enhancement efficiency (i.e., $B\rm_{w}$) is set to
1.6 and 10000, respectively. These models are labelled by
I, II and III in Table 1.  Excluding $q$ =1.6, we also adopted other two values
of mass ratio in the study (i.e.,  $q$=1.2 and 2.4) for metallicity $Z$=0.001.
These two models are labelled by IV and V in Table 1.
Moreover, other two values of $B\rm_{w}$ i.e., 5000 and 1000) were used in this
study for the models with $Z$=0.001 and $q$=1.6 to investigate its effects
on our final results, and these models are labelled  by VI and VII in Table 1.
For each model, four values of
initial helium abundance were used, i.e., $Y$=0.24, 0.28, 0.32 and 0.40.
There are a total of 28 sets of binary models combined with different input
parameters from model I to VII.

In tidally enhanced stellar-wind model (Lei et al. 2013a, b; also see
Han, Chen \& Lei 2010; Han et al. 2012; Han \& Lei 2014),
the mass-loss of primary star in a binary system is
determined by the initial orbital period.
Thus, the primary stars in binary systems
with long orbital period may locate on
RHB position due to little mass loss on
the RGB stage, while primary stars in binary
systems with shorter periods would lose
much envelope masses and settle on hotter HB positions
or even fail to ignite helium in their cores and
become a helium WD finally. For each set of
binary model list in Table 1, we
adopted different initial orbital periods in binary evolution
to find out  what kind of initial periods in binary would
produce BHk stars. From columns 4 to 7 of  Table 1, we  list
four critical periods for each binary model that
are important in our calculations.
Specifically, the initial orbital periods listed in column 4
(e.g., labelled by $P_{1}$) represent the minimum
periods for primary stars to experience normal helium
core flash at RGB tip. If the initial orbital period is
shorter than this one, the primary star would lose
much envelope mass on the RGB and experience early or late
hot flash before locating on HB. On the other
hand, initial periods listed in column 5
(e.g., labelled by $P_{2}$) for each set of model
present the minimum periods for primary stars to
undergo early hot flash, while the periods
labeled by $P_{3}$ in column 6 denote the minimum periods
for the primary stars to experience late hot flash.
If the initial orbital period of a binary is shorter than $P_{3}$
(e.g., periods list in the last column of Table 1, labeled by $P_{4}$),
the primary star may lose too much envelope mass on the RGB
and fail to ignite helium in its core, then dies as a
helium WD (see Fig. 1).

 In Table 1, for all the binary models with the highest helium abundance of $Y$=0.40
(e.g., set 4, 8, 12, 16, 20, 24 and 28), the minimum orbital periods for
primary stars to experience normal helium core flash at the RGB tip
(i.e., the orbital periods labeled by $P_{1}$ in Table 1)
are set to 10 000 d. This is because these models
present the highest helium abundance, thus the lowest
stellar mass at ZAMS. Therefore, even given a very long initial orbital period,
e.g., $P$=10 000 d, for which the two components in these binaries could be considered
 as two single stars, the primary star still  experience an early hot flash instead of a normal helium flash at
 the RGB tip [see panel (m) in Fig. 1]. These stars would become hot EHB stars rather than
 BHk stars after helium core flash, and it is beyond the range of this study.

\section{results}

In the tidally enhanced stellar-wind model,
we evolve the primary star from ZAMS to WD,
and obtain the evolution parameters (e.g., effective
temperatures, gravities, luminosity and surface
chemical composition etc.). These parameters
could be compared with observations directly or
with observations after translation.

\begin{figure}
\centering
\includegraphics[width=90mm]{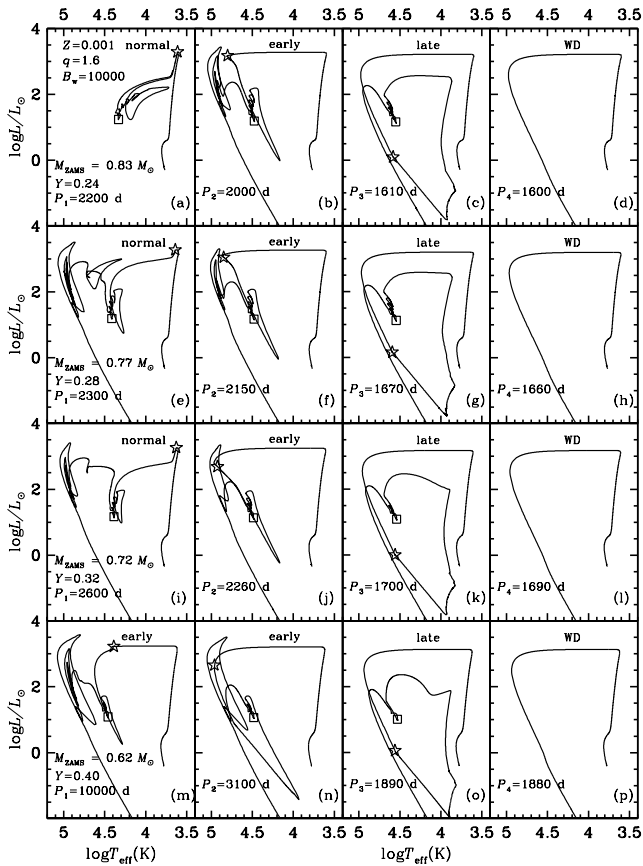}
\begin{minipage}[]{80mm}
\caption{The evolution tracks for the primary stars for binaries list in model II
of Table 1 (e.g., sets 5-8).  The
pentacles in each panel represent the positions where
the primary helium core flash takes place, while the
open squares denote ZAHB positions for each track. The initial
orbital periods for each binary system are labelled in each panel.  }
\end{minipage}
\end{figure}

\subsection{Evolution tracks}
 Fig. 1 shows the evolution tracks of primary stars from ZAMS to WD
for the binary systems of model II list in Table 1 (e.g., sets 5-8). Note that
the evolution tracks for other models list in Table 1 are not presented,
because they are very similar with the ones we showed here.
Specifically, panels (a)-(d) in Fig. 1 show the evolution tracks
for binary models with different initial orbital periods labelled by set 5 in Table 1
(i.e., $M_{\rm ZAMS}$=0.83$M_{\odot}$; $Y$=0.24), panels (e)-(h)
show the evolution tracks of set 6
(i.e., $M_{\rm ZAMS}$=0.77$M_{\odot}$; $Y$=0.28),
panels (i)-(l) present evolution tracks of set 7
(i.e., $M_{\rm ZAMS}$=0.72$M_{\odot}$; $Y$=0.32), while
panels (m)-(p) present evolution tracks of set 8
(i.e., $M_{\rm ZAMS}$=0.62$M_{\odot}$; $Y$=0.40).
The initial orbital periods
are also labeled in each panel. In Fig. 1, pentacles denote
the positions on H-R diagram where helium core flash
is taking place, while open squares denote the ZAHB positions
for each model. The label 'normal' in
Fig. 1 means the primary star experience normal
helium core flash at the RGB tip; 'early' means
the primary star experience helium core flash
on the way towards WD (i.e., early hot flash);
'late' means that helium core flash is taking place
on the WD cooling curve (i.e., late hot flash);
while 'WD' means the primary
stars lose too much envelope masses on the RGB stage to
ignite helium burning and become helium WDs finally.

As we discussed in Paper I (also see Brown et al. 2001, 2010, 2012),
early hot flash process could produce hot EHB stars
rather than BHk stars (also see Paper I), since the internal convection
mixing triggered by helium core flash can not
reach the hydrogen-enriched surface and change the
surface chemical composition. On the other hand,
due to the very thin hydrogen-enriched envelope,
convection mixing in late hot flash process could penetrate the surface
and engulf hydrogen into hotter interior.
Then, a helium and carbon-enhanced envelope is left, which
makes a star present higher temperature and faint luminosity
than normal EHB stars when settling on ZAHB.
As we described in Section 2, the initial orbital
periods labelled by $P_{2}$ in Table 1 denote the minimum periods
for early hot flash, below which the primary star
would experience late hot flash due to losing much
envelope mass, while initial orbital periods
labelled by $P_{3}$ in Table 1 represent the minimum
periods for late hot flash, below which the primary star
would fail to ignite helium burning in its core and
die as a helium WD due to losing too much envelope mass.
It means that, for each set of binary model list in Table 1,
the primary stars would experience late hot helium flash and
become BHk stars if the initial orbital
period is between $P_{2}$ and $P_{3}$ (i.e., $P_{3}$ $\leq$ $P$ $<$ $P_{2}$).
One can see from Fig. 1 that, even though their initial
helium abundances are very different,  all sets of binary models
could produce BHk stars within different period range
(e.g., 1610 $\leq$ $P$ $<$ 2000 for $Y$=0.24;
1670 $\leq$ $P$ $<$ 2150 for $Y$=0.28;
1700 $\leq$ $P$ $<$ 2260 for $Y$=0.32;
1890 $\leq$ $P$ $<$ 3100 for $Y$=0.40).
 The ranges of orbital periods to produce BHk stars
for other binary models with different input parameters
are shown in Table 1.
We will discuss the effects of the input
parameters on our results in Section 4, including metallicity,
mass ratio, tidal enhancement efficiency and helium abundance.

\begin{table*}
\small
 \begin{minipage}{140mm}
  \caption{The evolution parameters for the primary stars of model II in Table 1 when settling on
  ZAHB. From the left to the right of the table, it presents the initial orbital
  periods (in days), stellar mass at ZAHB ($M_{\rm ZAHB}$), effective temperature at
  ZAHB, luminosity at ZAHB,  gravity at ZAHB, the surface hydrogen, helium and carbon at
  ZAHB, and flash status. Flash status means what kind of helium flash does the
  primary star experience before settling on ZAHB. See the text for details.}
  \end{minipage}\\

    \begin{tabularx}{15.6cm}{cccccccccX}
\hline\noalign{\smallskip}
$P$/{\rm d} &$M_{\rm ZAHB}/M_{\odot}$ &  $\rm{log}\it{T}\rm_{eff}$   & $\rm{log}\it L/L_{\odot}$  & $\rm{log}\it{g}$  & $X\rm_{surf}$(by mass)  & $Y\rm_{surf}$(by mass) & $C\rm_{surf}$(by mass) & Flash status\\
\hline\noalign{\smallskip}
set 5  &$M_{\rm ZAMS}$=0.83$M_{\odot}$, &$Y$=0.24 \\
\hline\noalign{\smallskip}
2200 & 0.5036 & 4.3323 & 1.2393 & 5.1833 & 7.4431$\times10^{-1}$ & 2.5317$\times10^{-1}$ & 1.6144$\times10^{-4}$ &normal\\
2000 & 0.4829 & 4.4756 & 1.1941 & 5.7837 & 7.4431$\times10^{-1}$ & 2.5317$\times10^{-1}$ & 1.6161$\times10^{-4}$
&early\\
1610 & 0.4702 & 4.5498 & 1.1593 & 6.1035 & 1.3548$\times10^{-3}$ & 9.7006$\times10^{-1}$ & 1.0676$\times10^{-2}$
&late\\
\hline\noalign{\smallskip}
set 6  &$M_{\rm ZAMS}$=0.77$M_{\odot}$, &$Y$=0.28 \\
\hline\noalign{\smallskip}
2300 & 0.4822 & 4.4128 & 1.1881 & 5.5379 & 7.0702$\times10^{-1}$ & 2.9043$\times10^{-1}$ & 1.6325$\times10^{-4}$ &normal\\
2150 & 0.4752 & 4.4792 & 1.1696 & 5.8157 & 7.0702$\times10^{-1}$ & 2.9043$\times10^{-1}$ & 1.6325$\times10^{-4}$
&early\\
1670 & 0.4620 & 4.5454 & 1.1234 & 6.1140 & 1.5369$\times10^{-3}$ & 9.7103$\times10^{-1}$ & 1.0145$\times10^{-2}$
&late\\
\hline\noalign{\smallskip}
set 7  &$M_{\rm ZAMS}$=0.72$M_{\odot}$, &$Y$=0.32 \\
\hline\noalign{\smallskip}
2600 & 0.4811 & 4.3869 & 1.1770 & 5.4444 & 6.6936$\times10^{-1}$ & 3.2820$\times10^{-1}$ & 1.6487$\times10^{-4}$ &normal\\
2260 & 0.4673 & 4.4859 & 1.1384 & 5.8662 & 6.6936$\times10^{-1}$ & 3.2820$\times10^{-1}$ & 1.6487$\times10^{-4}$
&early\\
1700 & 0.4535 & 4.5401 & 1.0881 & 6.1204 & 4.8556$\times10^{-3}$ & 9.6951$\times10^{-1}$ & 1.0017$\times10^{-2}$
&late\\
\hline\noalign{\smallskip}
set 8  &$M_{\rm ZAMS}$=0.62$M_{\odot}$, &$Y$=0.40 \\
\hline\noalign{\smallskip}
10000 &0.4559 & 4.4587 & 1.0871 & 5.7981 & 5.9338$\times10^{-1}$ & 4.0440$\times10^{-1}$ & 1.7109$\times10^{-4}$
&early\\
3100 & 0.4503 & 4.4821 & 1.0635 & 5.9099 & 5.9338$\times10^{-1}$ & 4.0440$\times10^{-1}$ & 1.7109$\times10^{-4}$
&early\\
1890 & 0.4365 & 4.5293 & 1.0101 & 6.1385 & 1.0158$\times10^{-2}$ & 9.6598$\times10^{-1}$ & 1.1008$\times10^{-2}$
&late\\
\hline\noalign{\smallskip}
  \end{tabularx}
\end{table*}

\subsection{Evolution parameters at  ZAHB}
We also presented the  evolution parameters at ZAHB for the primary stars
presented in Fig. 1. From the left to the right of Table 2, it gives
initial orbital period (in days), stellar mass at
ZAHB ($M_{\rm ZAHB}$), effective temperature, gravity,
surface hydrogen, surface helium, surface carbon and flash status, respectively.
Flash status in Table 2 denote what kind of helium core flash
does the primary star experience before locating on ZAHB,
and the labels 'normal', 'early' and
'late' have  the same meaning as we described in Fig. 1.
The evolution parameters for the tracks that have failed to
ignite helium burning in their cores are not listed in Table 2.

As we discussed in Paper I, the surface chemical abundances
are not changed by internal flash mixing in the
'normal' and 'early' models.
However, due to losing more envelope mass
on the RGB stage, 'early' models present higher effective
temperatures than 'normal' models by about several thousands of K at ZAHB.
On the other hand, 'late' models experience late
hot flash when descending WD cooling curve. They
present the highest effective temperatures and
helium and carbon enhancement in the surface due
to internal flash mixing. One can see from Table 2,
with initial helium abundance increasing from $Y$=0.24
to 0.40, that the effective temperatures and luminosity
of 'late' models in each set model decrease slowly, while their
gravities increase gradually. This is because that, for the 'late' models,
core helium burning provide the mainly energies released
at HB stage (e.g., the hydrogen-enriched envelope is too
thin to ignite hydrogen burning during HB stage; Brown et al. 2001;
Cassisi et al. 2003). However, for
higher initial helium abundance, the stellar masses
at ZAMS ($M_{\rm ZAMS}$) are smaller if stars have same ages, thus these stars
have smaller helium core mass when settling on ZAHB, and
present lower temperatures and luminosity  but larger gravities
than the stars with lower initial helium abundance.

 One can also see from Table 2 that the 'normal' and 'early' models show
normal hydrogen, helium and carbon abundance in the surface, because
the flash progress does not change the chemical composition in the surface.
However, for the 'late' models (e.g., the model with $P$=1610 d in set 5 of Table 2),
flash triggered mixing could enhance helium
up to 0.9701 and  carbon up to 0.0107 in the surface.
 Miller Bertolami et al. (2008) also studied the late hot flash scenario for the
 formation of He-rich subdwarf stars with a wide range of metallicities and
 physical assumptions. For their model with $Z$=0.001, they found a helium
 enhancement in the surface up to 0.9666 and carbon enhancement up to
 0.0107. Their results are very similar with  our results found here.  Furthermore,
 Miller Bertolami also found a transition type of late hot flash which is called the
 'shallow' mixing cases (also see Lanz et al. 2004). In this kind of flash,  the hydrogen
 envelope is mixed only with the convective shell in the outer part of the core, thus
 none of the hydrogen captured by the convective shell will be burned.  Therefore,
 the surface helium abundance is lower than the case that experience 'deep' mixing (e.g.,
 $Y$=0.7475 in the 'shallow' model with $Z$=0.001 listed in table 3 in Miller Bertolami et al. 2008 ).
 However, this kind of hot  flash is rare for low metallicities (Lanz et al. 2004; Miller Betrolami et al. 2008).
 Latour et al. (2014) obtained the fundamental parameters (e.g., effective temperature,
 gravity, helium and carbon abundance) for 38 hot subdwarf stars in $\omega$ Cen.
 They found that,  for some He-rich hot subdwarf stars, the surface helium enhancement could reach 0.92 and
 carbon enhancement could reach 0.015. Their results are consistent with ours presented here.
 Moreover Latour et al. (2014) also found a positive correlation between helium and carbon enhancement
 among He-rich subdwarf stars in their sample, which support the late hot flash scenario for the
 formation of BHk stars in $\omega$ Cen.

\subsection{Comparison with observation in $\it{T}\rm_{eff}$-$\rm{log}\it{g}$ diagram}

\begin{figure}
\centering
\includegraphics[width=89mm]{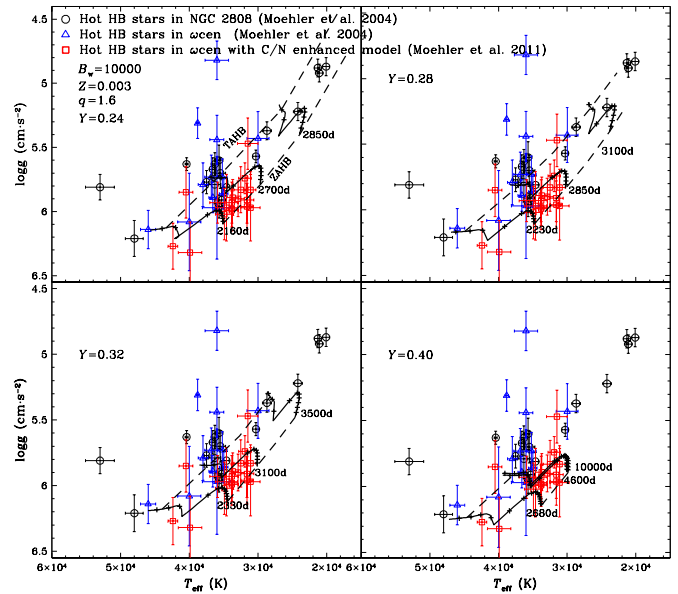}
\begin{minipage}[]{90mm}
\caption{Comparison between the results of model I   and observations in the
$\it{T}\rm_{eff}$-$\rm{log}\it{g}$ plane of NGC 2808 and $\omega$ Cen .  The
input parameters adopted in model I are labeled,
and the time interval between two adjacent + symbols in each track is $10^{7}$~yr}
\end{minipage}
\end{figure}

 BHk stars were first found in massive GC NGC 2808 and $\omega$ Cen
(Whitney et al. 1998; D'Cruz et al. 2000).  NGC 2808 is a  typical GC that
hosts splitting  MS and extended HB (Piotto et al. 2002, 2007).  Its metallicity
is about $Z$=0.0014 or [Fe/H]=-1.15 (Harris 1996, version 2003) and its age  is in the range of
10.4-12.9 Gyr (Gratton et al. 2010).  $\omega$ Cen is a complex system among GCs,
and some researchers considered it as the surviving nucleus of a dwarf galaxy captured and
 disrupted by Milky Way several Gyr ago (Dinescu et al. 1999; Bellazzini et al. 2008; Marconi et al. 2014).
 It presents two parallel MS and at least five stellar populations in sub giant branch
 ( Bedin et al. 2004; Piotto et al. 2005; Villanova et al. 2007, 2014). Furthermore,
$\omega$ Cen cover a wide range of metallicity, e.g., -2.2$\leq$ [Fe/H]$\leq$-0.6
(Johnson \& Pilachowski 2010; Villanova et al. 2014).
Since the parameters of NGC 2808 are very similar with the values we
adopted here, we compared our results with this GC in the $\it{T}\rm_{eff}$-$\rm{log}\it{g}$ diagram.
Though $\omega$ Cen cover a wide range of metallicity,
we  compared our results with this GC in $\it{T}\rm_{eff}$-$\rm{log}\it{g}$ diagram as well,
because we also adopted three different values of metallicity in
our calculations, e.g., 0.003 or [Fe/H]=-0.8, 0.001 or [Fe/H]=-1.3 and 0.0003 or [Fe/H]=-1.8, and
these values are in the metallicity range of $\omega$ Cen.

 Figs 2-8 present the comparison between our model calculation results
(i.e., evolution tracks at HB stage from models I to VII listed in Table 1) and observation in
$\it{T}\rm_{eff}$-$\rm{log}\it{g}$ diagram of
GC NGC 2808 and $\omega$ Cen.
The black open circles in these figures represent BHk stars and
hot EHB stars in GC NGC 2808, and the blue open triangles represent
BHk stars and hot EHB stars in $\omega$ Cen. These spectral
parameter data for the hot HB stars in NGC 2808 and $\omega$ Cen are from
Moehler et al. (2004),   while the red open squares in these figures
are the hot He-rich HB stars in $\omega$ Cen studied by Moehler et al. (2011).
The spectral parameters of these hot stars are obtained by using C/N-enhanced model
atmospheres with mass fraction of 3\% and 1\% for carbon and nitrogen, which is predicted
by the late hot flash scenario (Lanz et al. 2004; Miller Bertolami et al. 2008).
When C/N-enhanced model atmospheres are used, the obtained gravities of the hot HB stars
would be higher than the values obtained without C/N-enhanced model atmospheres
by an average of about 0.1 dex (see table 4 in Moehler et al. 2011).
The dashed line at the bottom labeled by 'ZAHB' in each panel of Figs 2-8
represent the ZAHB positions for our
evolution tracks, while the dashed line at the top
labelled by 'TAHB' represent the terminal age HB (TAHB), which
is defined as the stage when central helium abundance dropped
below $1\times10^{-4}$.
The black solid curves in each panel of Fig. 2 to 8 denote
the evolution tracks at HB stage for the primary stars
listed in table 1. The value of initial orbital
periods (in days) are also labeled nearby the tracks, and
the input parameters for these models are also labelled.

 Fig 2 shows the comparison between the  results of model I
list in Table 1 and the observation.  These models have the
input parameters of $Z$=0.003, $q$=1.6 and $B\rm_{w}$=10000.
The initial helium of the binary models from upper left to bottom right panel is $Y$=0.24,
0.28, 0.32 and 0.40, respectively.
One can see that in the upper left panel (i.e., $Y$=0.24), most of the BHk stars
(e.g., $T\rm_{eff} > 32000$ K; Moni bidin et al. 2012)
in NGC 2808 and $\omega$ Cen from Moehler et al. (2004)  locate
in the region predicted by our models.
However, the gravities of BHk stars predicted by
our models seem to be a little larger than the values obtained
by Moehler et al. (2004). This discrepancy between model calculations
and observations becomes more evident when initial helium abundance
becomes higher (see other three panels in fig 2
with $Y$=0.28, 0.32, 0.40). Especially for $Y$=0.40 (bottom right panel in fig 2), one can see
that most of the BHk stars from Moehler et al. (2004),
e.g., black open circles and black open triangles,
locate out of the HB region defined by
ZAHB and TAHB line. A possible
explanation for this discrepancy between model calculations and
observations would be that the late hot flash scenario predicted
carbon and nitrogen enhancement in the surface of BHk stars (e.g.,
$3\%$ and $1\%$ for carbon and nitrogen by mass in Lanz et al. 2004).
However, this enhancement was not considered in the atmosphere model
used to obtain the atmosphere parameters of BHk stars in NGC 2808 and
$\omega$ Cen in Moehler et al. (2004).
On the other hand, Moehler et al. (2011) did consider
the carbon and nitrogen enhancement in
the atmosphere model when obtaining
the parameters of some BHk stars in $\omega$ Cen.
These BHk stars are denoted by red open squares in
figs 2-8.
One can see clearly in fig 2 that when C/N-enhanced model atmospheres
are considered, most of the BHk sample stars in Moehler et al. (2011) are well predicted by
our late hot flash models, regardless of the helium abundance adopted, including the highest value of
$Y$=0.40.  This is also the case for figs 3-8.
Thus, if carbon and nitrogen enhancement is considered, one would expect a better agreement between
our calculation results and the BHk stars in NGC 2808 and $\omega$ Cen from
Moehler et al. (2004).

 One can see from figs 2-8,  that under the tidally enhanced stellar wind,
all the binary models from I to VII could
produce BHk stars, though the input parameters are
different with each other. For
figs 2-4, which correspond to models I , II and III in Table 1,
the binaries in these models have the same mass ratio of primary to
secondary (i.e., $q$=1.6) and tidal enhancement efficiency (e.g., $B\rm_{w}$=10000), but
with different metallicities (e.g., $Z$=0.003, 001 and 0.0003 for model I, II and III, respectively);
figs 3, 5 and 6, which correspond to  models II, IV and V, present
the same metallicity (i.e., $Z$=0.001) and tidal enhancement efficiency
(i.e., $B\rm_{w}$=10000), but with different mass ratios (e.g., $q$=1.6, 1.2 and 2.4
for models II, IV and V respectively). While figs 3, 7 and 8, which correspond to
models II, VI and VII, have the same metallicity (i.e., $Z$=0.001) and
mass ratio (i.e., $q$=1.6), but with different tidal enhancement efficiency
(e.g., $B\rm_{w}$=10000, 1000 and 5000 respectively).
We will discuss the effects of the various input parameters on
our results in next section.

\begin{figure}
\centering
\includegraphics[width=89mm]{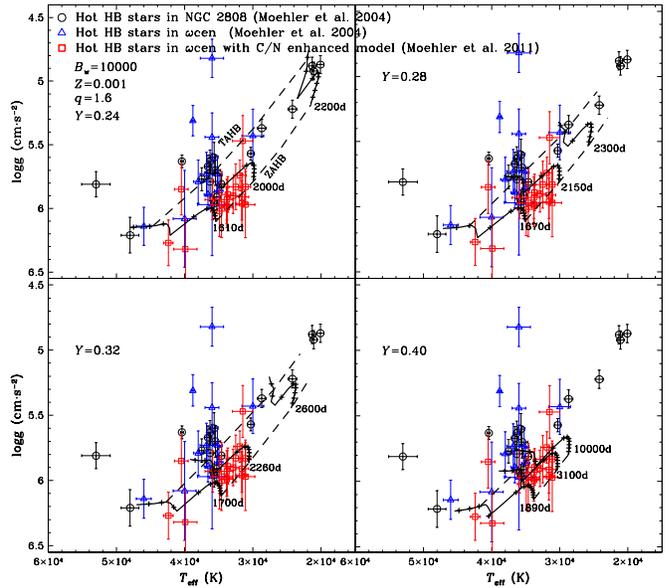}
\begin{minipage}[]{90mm}
\caption{The same as Fig. 2, but for the comparison between observations and model II. }
\end{minipage}
\end{figure}

\begin{figure}
\centering
\includegraphics[width=89mm]{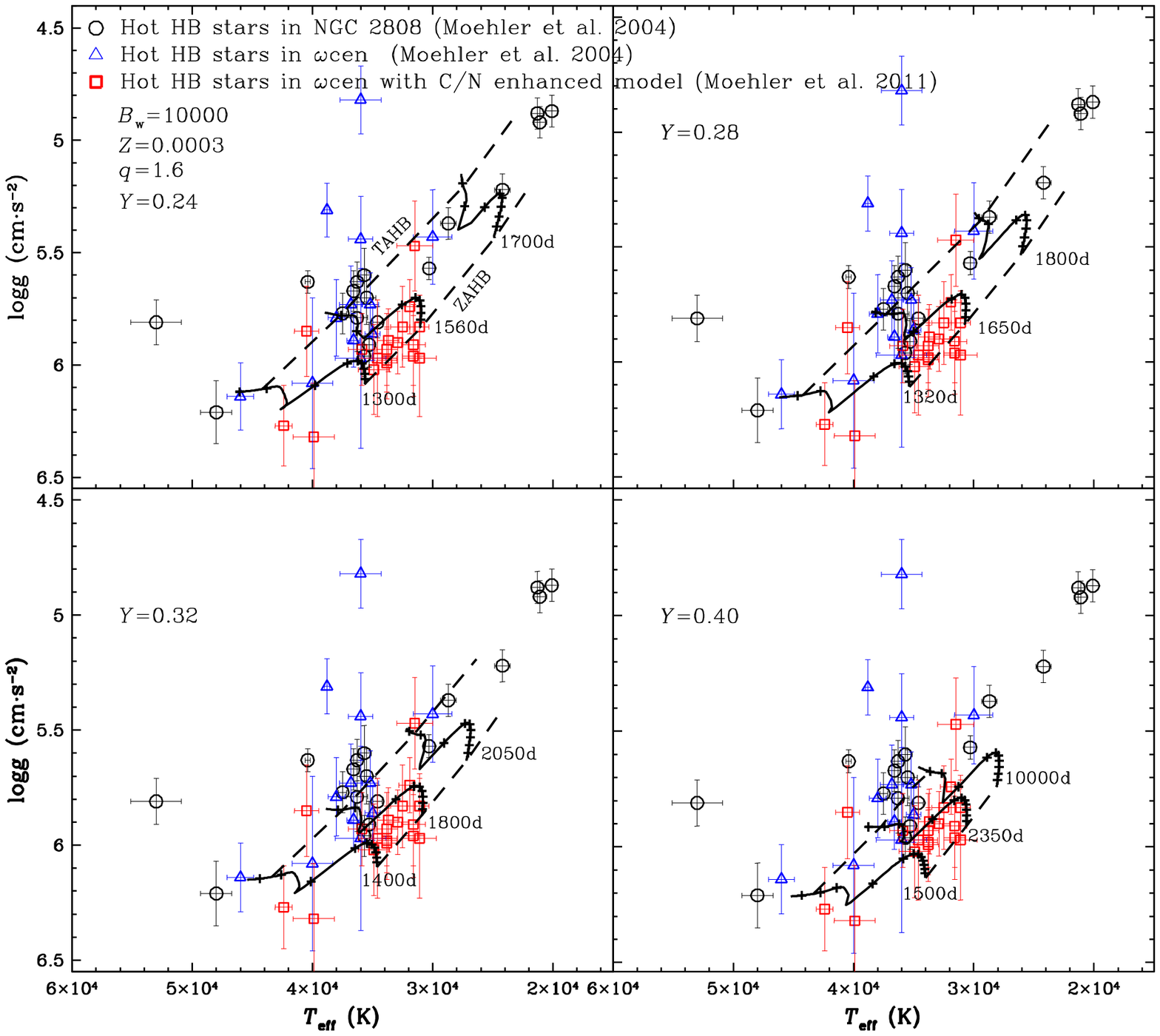}
\begin{minipage}[]{90mm}
\caption{The same as Fig. 2, but for the comparison between observations and model III. }
\end{minipage}
\end{figure}

\begin{figure}
\centering
\includegraphics[width=89mm]{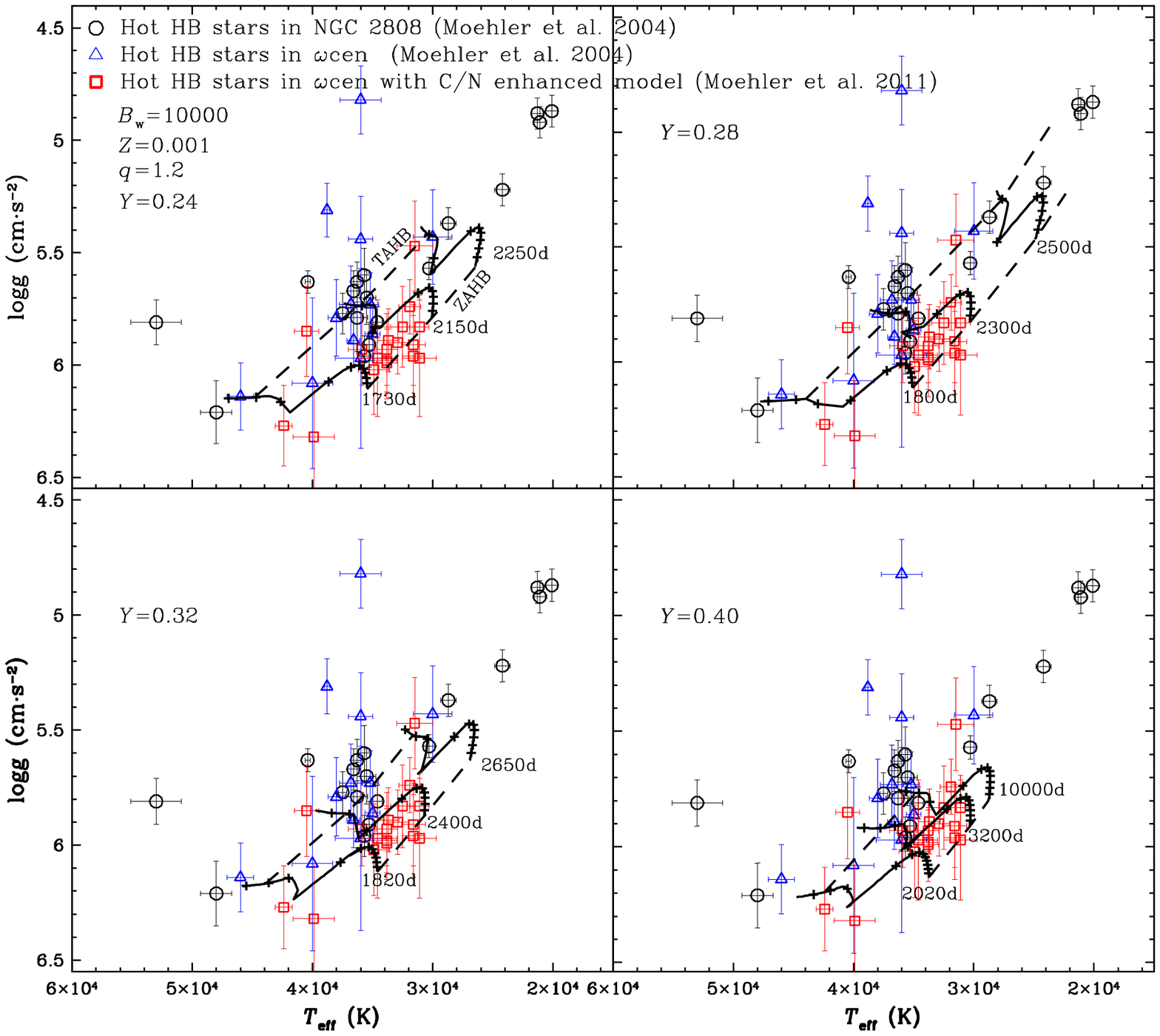}
\begin{minipage}[]{90mm}
\caption{The same as Fig. 2, but for the comparison between observations and model IV. }
\end{minipage}
\end{figure}

\begin{figure}
\centering
\includegraphics[width=89mm]{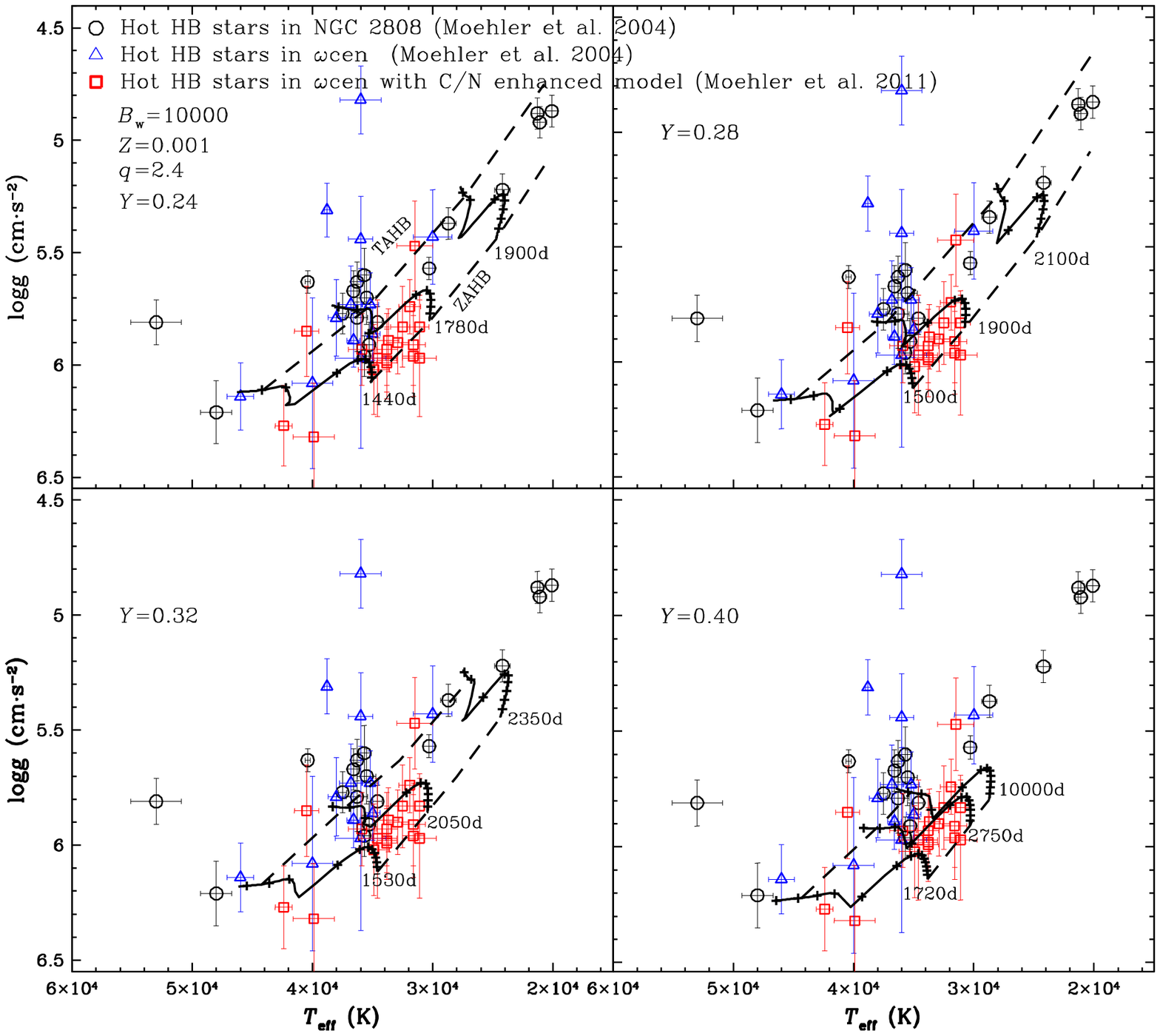}
\begin{minipage}[]{90mm}
\caption{The same as Fig. 2, but for the comparison between observations and model V. }
\end{minipage}
\end{figure}

\begin{figure}
\centering
\includegraphics[width=89mm]{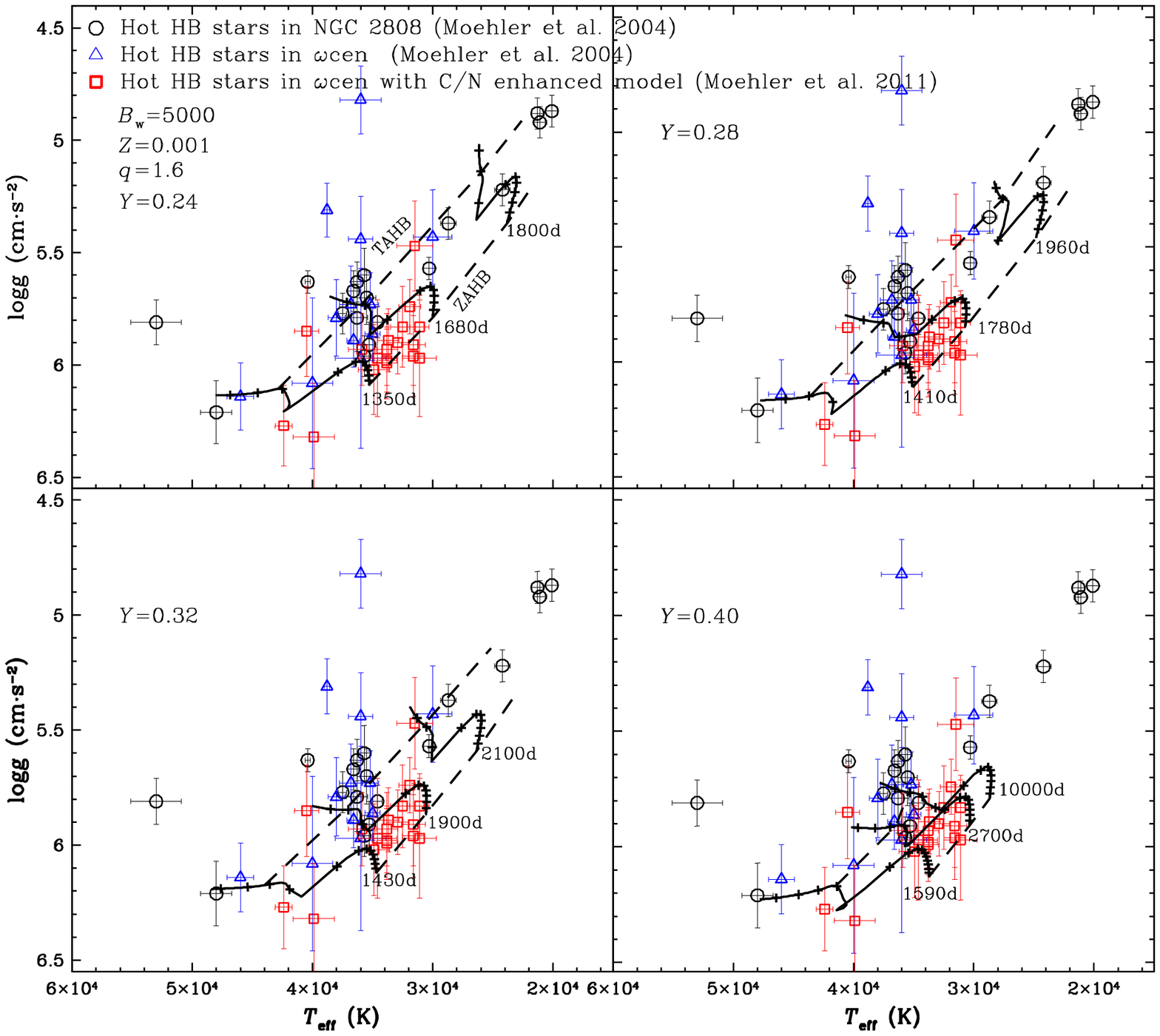}
\begin{minipage}[]{90mm}
\caption{The same as Fig. 2, but for the comparison between observations and model VI. }
\end{minipage}
\end{figure}

\begin{figure}
\centering
\includegraphics[width=89mm]{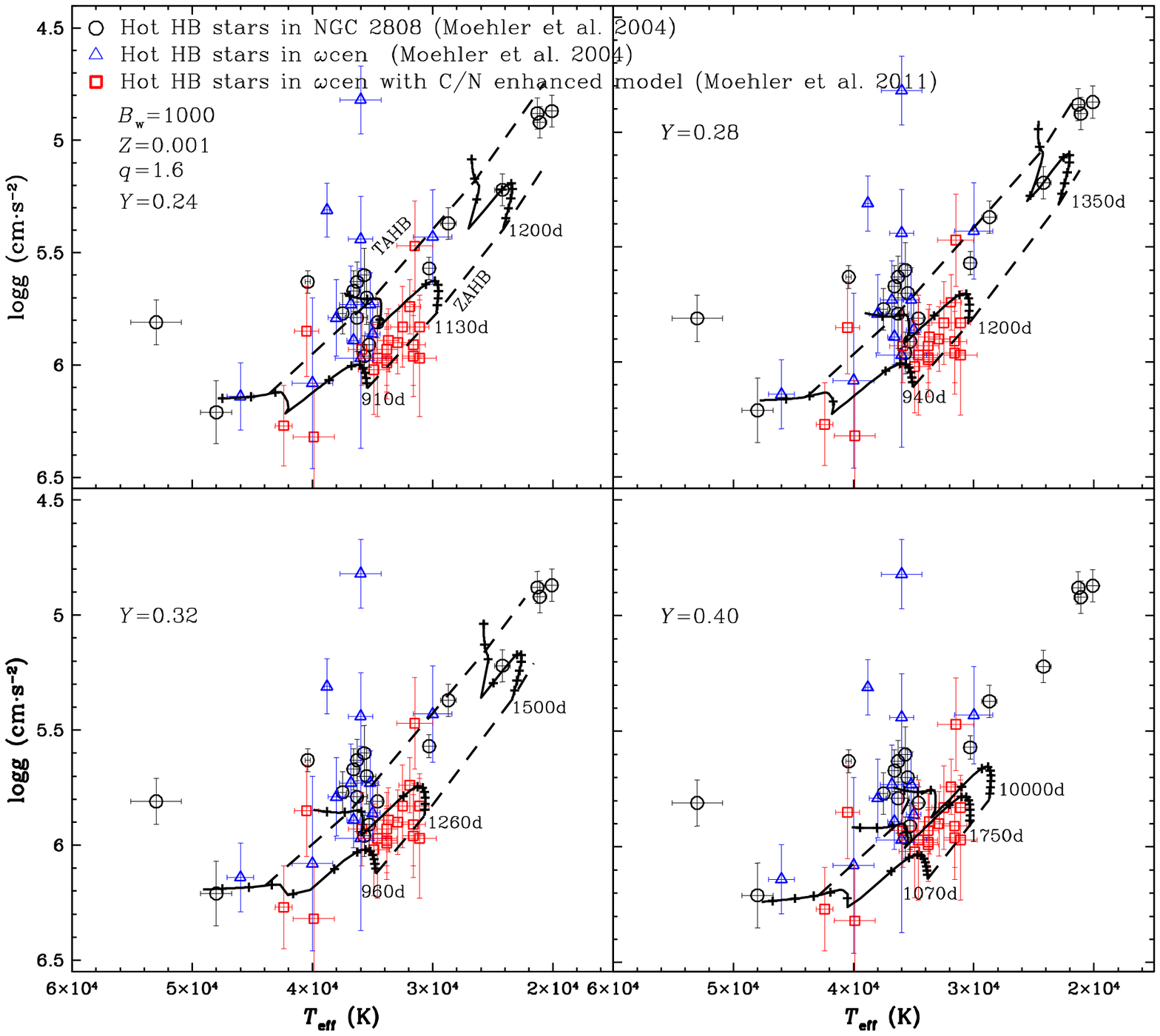}
\begin{minipage}[]{90mm}
\caption{The same as Fig. 2, but for the comparison between observations and model VII. }
\end{minipage}
\end{figure}

\begin{figure}
\centering
\includegraphics[width=80mm]{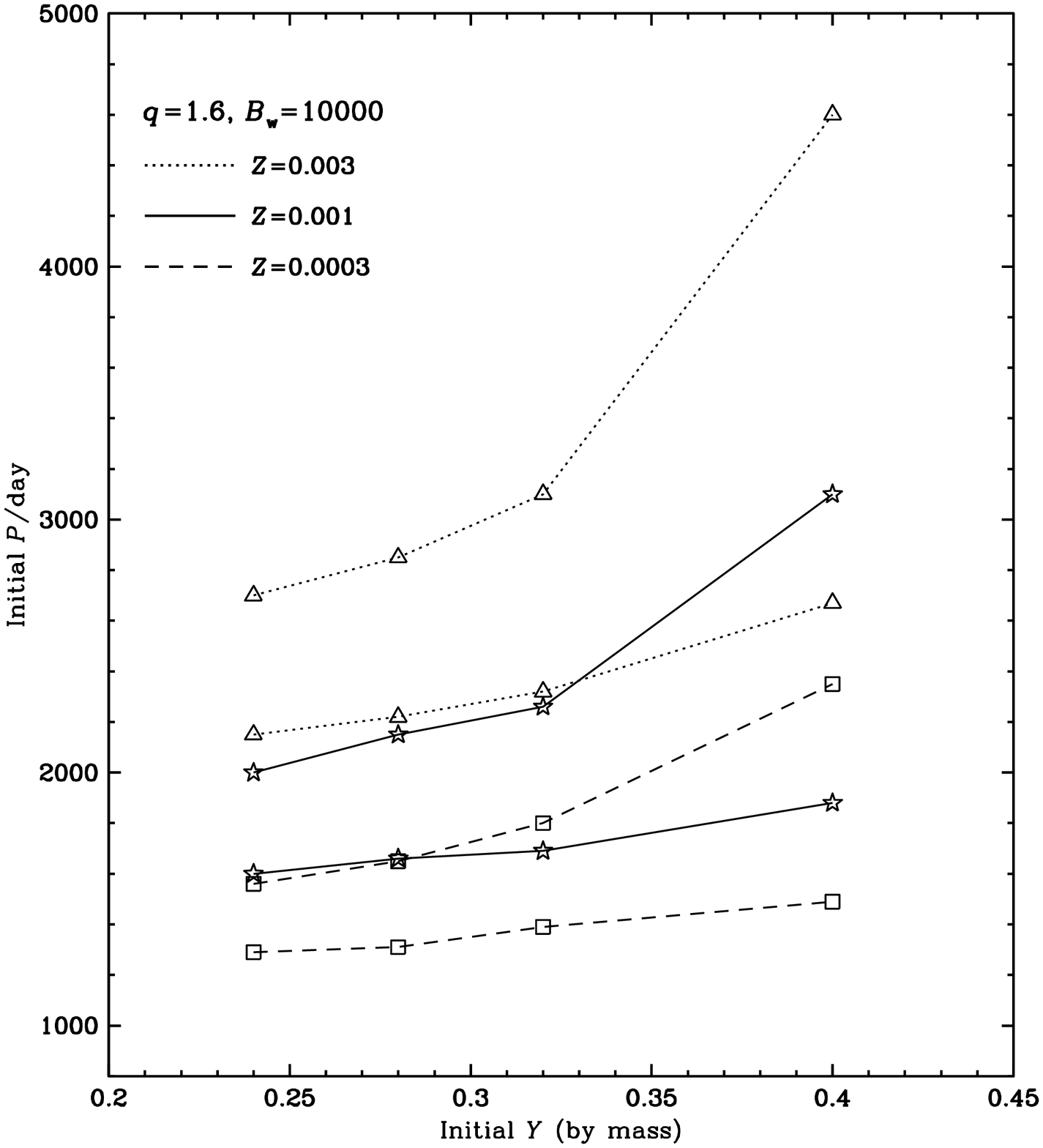}
\begin{minipage}[]{85mm}
\caption{The correlation between the range of initial orbital periods to produce BHk stars with
increasing helium abundance for different metallicities. }
\end{minipage}
\end{figure}

\begin{figure}
\centering
\includegraphics[width=80mm]{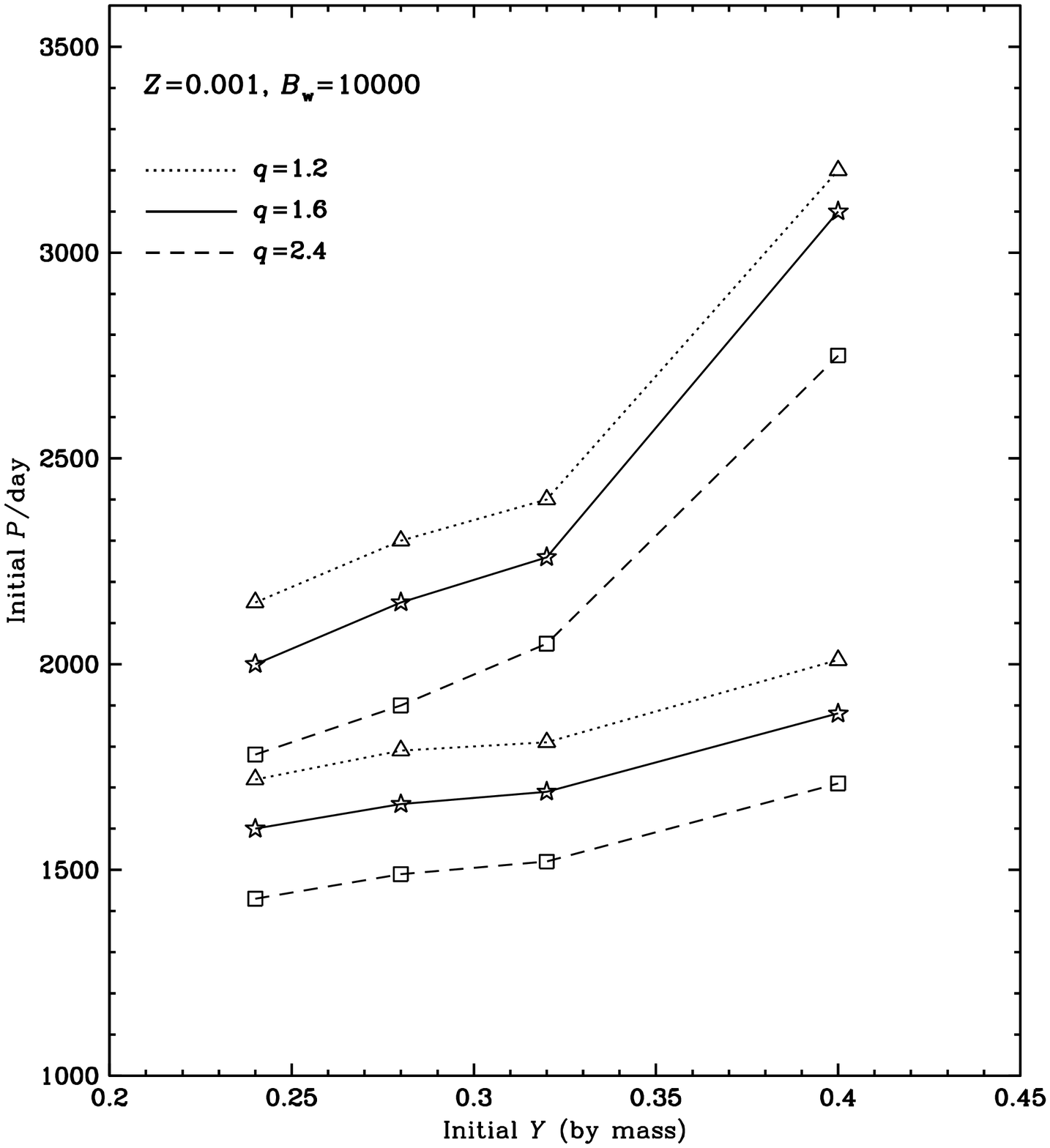}
\begin{minipage}[]{85mm}
\caption{The correlation between the range of initial orbital periods to produce BHk stars with
increasing helium abundance for different mass ratios. }
\end{minipage}
\end{figure}

\begin{figure}
\centering
\includegraphics[width=80mm]{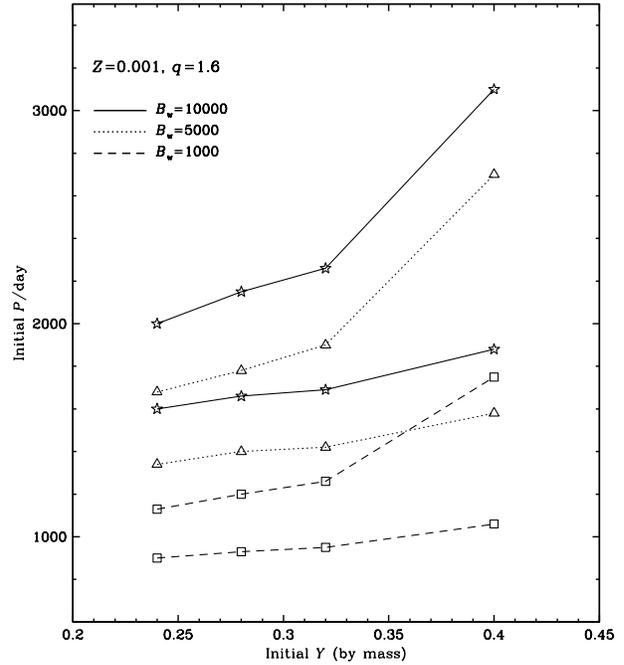}
\begin{minipage}[]{85mm}
\caption{The correlation between the range of initial orbital periods to produce BHk stars with
increasing helium abundance for different tidal enhancement efficiency. }
\end{minipage}
\end{figure}

\section{discussion}

 In the tidally enhanced stellar-wind model, the primary
star would lose huge envelope mass during RGB stage and experience
late hot helium flash on the WD cooling curve to become a
BHk star. Though the mass loss of the primary stars in this
scenario depends on the initial orbital periods of the binaries
(Paper I, also see the discussion in Section 2 of this study),
other input parameters such as metallicity, mass ratio of primary to
secondary, tidal enhancement efficiency and helium abundance, would
influence the binary evolution, thus influence the formation of BHk stars
in binaries.  Therefore, we discussed the effects of these input parameters
on our results in this section, and we also discussed our results with recent
observations and other theoretical models.

\subsection{Metallicity, $Z$}
Fig 9 shows relationship  between the range of initial orbital
periods which is needed to produce BHk stars and
the initial helium abundance for models I-III.
These models have the same $q$ and $B\rm_{w}$ (i.e., $q$=1.6, $B\rm_{w}$=10000),
but different metallicities (e.g., $Z$=0.003, 0.001 and 0.0003 for models I, II and III respectively. )
The top dotted line in Fig. 9 represents the minimum
orbital periods for the primary stars in model I to experience early hot flash (labeled by
$P_{2}$ in Table 1), below which
the primary stars would experience late hot flash and become BHk stars
(it also can be considered as the maximum period to produced BHk stars, see Section 3.1).
While the bottom dotted line denotes the minimum orbital periods for
the primary stars in model I to experience late hot flash and become BHk
stars (labelled by $P_{3}$ in Table 1), below which the primary stars would
lose too much envelope mass and die as helium WDs.
Therefore, the region between the two dotted lines in Fig. 9
denotes the range of orbital periods in which the binaries would
produce BHk stars for model I.  Similarly, the region marked by
two solid lines in Fig. 9 shows the range of orbital periods for the
binaries in model II to produce BHk stars, while the region
between two dashed lines presents the range of orbital
periods for the binaries in model III to produce BHk stars.

 One  can see from fig 9 that, with the metallicity increasing from
model III to model I, the range of orbital period to produce BHk stars
becomes wider. This trend is more evident at $Y$=0.40 for
these three models, e.g., $P_{2}-P_{3}  \approx $800 d  for model III ($Z$=0.0003) but
increases to about 2000 d for model I ($Z$=0.003).
This is because that higher metallicity would lead to
larger stellar wind on the RGB stage. Thus,
a longer orbital period in binary also could lose
enough envelope mass on the RGB stage to experience
late hot flash and become BHk stars.  It means that,
in the tidally-enhanced-stellar-wind scenario, GCs with higher metallicities
would produce BHk stars more easily than metal-poor ones if  a flat
distribution of the orbital periods for binaries in GCs is expected.
However, the detailed orbital periods for binaries
and their distributions in GCs  are difficult to obtained (Milone et al. 2012, 2016).
Massive GC NGC 6388 and NGC 6441 are found to host
extended HB morphologies and BHk stars (Rich et al. 1997;
Busso et al. 2007) that would  be  hardly
expected in the GCs with higher metallicity (e.g., [Fe/H]=-0.60
and -0.53 for NGC 6388 and NGC 6441 respectively; Harris 1996, version 2003).
Therefore, our results obtained here would be a possible explanation
for the formation of BHk stars in metal-enriched GC NGC 6388 and NGC 6441.
However, a further study for the effects of binaries on the formation of
BHk stars in these two GCs is needed, including $N$-body simulation to calculate the
distribution of orbital periods for the binaries in these two GCs.
For model I in fig 9 (i.e., $Z$=0.003), with the
initial helium abundance increasing, the orbital period range
to produce BHk stars becomes larger (e.g., $P_{2}-P_{3}  \approx $500 d  for $Y$=0.24, but
this value is about 2000 d for $Y$=0.40).   This
trend is also presented in models II and III in Fig. 9 that
have initial metallicity of $Z$=0.001 and 0.0003, respectively.
We will discuss this result in Section 4.4.

\subsection{Mass ratio, $q$}
 To study the effects of mass ratio on our results, we
showed in Fig 10 the relationship between the range of orbital periods
that is needed to produced BHk stars and initial helium abundance
for models II, IV and V listed in Table 1. These three models
have the same metallicity ($Z$=0.001) and tidal enhancement efficiency
($B\rm_{w}$=10000), but with different mass ratios (e.g., $q$=1.6, 1.2 and 2.4 for
models II, IV and V, respectively). As we described in
Fig. 9, the region defined by the two lines with same type
denotes the range of orbital periods in which BHk stars could be produced
in binaries. Specifically, the top lines in each pair represents the maximum
orbital periods to produce BHk stars, while the bottom lines in each pair
represents the minimum orbital periods to produce BHk stars.
From top to bottom of fig 10, the region defined by dotted lines
is the range of orbital periods to produce BHk star for binaries in
model IV ($q$=1.2), and the region defined by solid lines are is
the orbital periods range to produce BHk stars for the binaries in model
II ($q$=1.6), while the two dashed lines define the orbital
periods range for the binaries in model V ($q$=2.4) to produce
BHk stars.

One can see from Fig. 10 that the ranges of orbital periods
to produce BHk stars are very similar from $Y$=0.24 to $Y$=0.40
among the three models with different mass ratios. It means that no evident
correlations between mass ratio and the range of orbital periods that is needed
to produce BHk stars in binaries are found. This is because of the fact that,
in tidally enhanced stellar-wind model, smaller  mass ratio (or more massive
secondary) would expect much mass-loss for the primary, thus the
orbital periods needed to produce BHk stars would be  longer than
the values for binaries with larger mass ratio. However, the
whole space range of orbital periods to produce BHk star is
not altered obviously. On the other hand, the space range of orbital
periods to produce BHk stars for each model becomes wider with
the increase of initial helium abundance, which is also found in
Fig 9.

\subsection{Tidal enhancement efficiency, $B\rm_{w}$}
 The tidal enhancement efficiency, $B\rm_{w}$, is a free
parameter in tidally-enhanced-stellar-wind model. Tout \& Eggleton
(1988) used a value of 10000 to explain the mass inversion phenomenon
found in some RS CVn type binaries (Popper \& Ulrich 1977; Popper 1980).
We adopted three values of $B\rm_{w}$ in our calculation to
study its effects on our final results. Fig. 11 shows the relationship
between the space range of orbital periods to produce BHk stars
and the initial helium abundance
for the binaries in models II, VI and VII. These three models have
the same metallicity ($Z$=0.001) and mass ratio ($q$=1.6), but with
different $B\rm_{w}$ (e.g., $B\rm_{w}$=10000, 5000 and 1000 for
models II, VI and VII, respectively; see Table 1). The lines in
pair with different types have the same meaning as we described in
figs 9 and 10. The two solid lines mark the space range of orbital periods
which is needed to produce BHk stars for the binaries in model II ($B\rm_{w}$=10000);
the two dotted lines define the space range of orbital periods to produce
BHk stars for the binaries in model VI ($B\rm_{w}$=5000), while
the space range marked by two dashed lines denotes the orbital periods
to produce BHk stars for model VII ($B\rm_{w}$=1000).

 As expected, for a smaller value of  $B\rm_{w}$,
the orbital periods needed for binaries to
produce BHk stars are  shorter than the ones needed
for the binaries with larger value of $B\rm_{w}$ to produce
BHk stars.  When the value of tidal enhancement
increasing from 1000 to 10000 in fig 11, the space range of
orbital periods for the binaries to produce BHk stars
would become a little larger (e.g., for $Y$= 0.40,
$P_{2}-P_{3}  \approx $700, 1100 and
1200 d for $B\rm_{w}$=1000, 5000 and 10000 respectively),
but this trend is not very evident at lower $Y$, as we discussed in
Paper I.
We also found in fig. 11 that with the
initial helium abundance increasing in each model, the space range of
orbital periods for binaries to produce BHk stars becomes wider,
especially for the highest helium abundance, $Y$=0.40, which is the
same as we see in fig 9 and 10.

\subsection{Helium abundance, $Y$}
 For each model showed in figs 9-11, the
space range of orbital periods for binaries to produce
BHk stars becomes wider obviously with the increase in initial
helium abundance. This trend is very evident
for all the models regardless of the adopted metallicities, mass ratios  and
tidal enhancement efficiency.
This is due to the fact that stars with
higher helium abundance evolve faster
than normal stars. For a fixed age at the
RGB tip, helium enriched stars have smaller stellar
masses at ZAMS ($M_{\rm ZAMS}$) than normal stars.
Therefore, even in a wider binary system, these stars
still could experience late hot helium flash and become
BHk stars after losing a little envelope mass on the RGB stage, and
the range of initial orbital periods needed to produce BHk stars in binaries with higher
helium abundance is wider than the ones in binaries with lower helium abundance.
Our result presented here means that helium abundance would be
an important parameters in our binary models.
With a higher value of $Y$, one would expect that
BHk stars could be produced more easily in binaries if
a flat distribution of the orbital periods is assumed in
GCs. However, the detailed distribution of
orbital periods for binaries in GCs is difficult to obtain from
observation, but $N$-body simulation may be an alternative option.

Recently, multiple populations are found as a common
phenomenon among Galactic GCs, e.g., the splitting of MS, light
elements anomalies (Piotto et al. 2007; Gratton Carratta \& Bragaglia 2012; Gratton et al. 2013, 2014).
Helium enhancement due to self-enrichment in GCs is considered
to be correlated with this phenomenon (D'Antona et al. 2002, D'Antona \& Caloi 2008; Milone 2015).
When these heium enhanced second-generation  stars evolve into HB stage, they would
occupy BHB even EHB positions due to their higher
helium abundance, and some of them would become BHk stars (D'Antona et al. 2002; 2010).
These results seem to be consistent with the results we presented here such that
BHk stars would be produced more easily in binary systems with higher initial
helium abundance.

\subsection{Comparison with other models}
D'Antona et al. (2010) proposed that the BHk stars
would be the progeny of the blue MS in $\omega$ Cen , which could be
explained by using enhanced helium abundance of $Y$=0.38-0.40
(Moehler et al. 2007). They assumed that these blue MS stars
may experience extra mixing at upper RGB, which could increase the surface
helium abundance up to $Y\thickapprox$ 0.8. Then these stars evolve along a non-canonical evolution tracks
and settle on BHk regions when central helium ignites.
However, the helium enhanced model proposed by
D'Antona et al. (2010) could not predict other chemical anomalies in the
stellar surface, such as carbon and nitrogen. In fact,
using the MXU mode of the Focal Reducer and Spectrograph
of the Very Large Telescope, Latour et al. (2014) found a clear
positive relationship between helium and carbon enhancement
in some BHk stars of $\omega$ Cen, which is also predicted by our model
(see carbon and helium abundance of 'late' models listed in Table 2, also see the discussion in paper I.
Note that we do not consider chemical diffusion and gravitational settling
process in our models).
This observation evidence favors  late hot flash
scenario for the formation of BHk stars in GCs.

Nevertheless, the luminosity (or magnitude) extension
of BHk stars in H-R diagram (or CMD)
of some GCs is difficult to understand for most of the scenarios
explaining the formation of BHk stars (Brown et al. 2010; D'Antona et al. 2010;
Paper I),
including late hot flash scenario.
Recently, Tailo et al (2015) suggested that a rapidly rotating
second-generation progenitor for BHk stars in $\omega$ Cen would result
in a larger helium core mass by up to $\sim$0.04$M_{\odot}$ when helium core
flash is taking place, thus a higher luminosity for stars at ZAHB.
 They suggested that the pre-MS accretion disc of progenitors
for BHk stars may suffer an early disruption, which results in
a faster rotation for the following evolution and thus
an increase in helium core mass at helium flash.
Employing this mechanism, they could explain the luminosity range of
BHk stars in $\omega$ Cen.
Note that the results in Tailo et al (2015) would not
conflict with our results obtained in this study.
In Tailo et al (2015), the BHk progenitors also need to
experience huge mass loss on the RGB stage and undergo
late hot helium flash before becoming BHk stars. However,
they did not consider the physical mechanism of huge mass loss
on the RGB in details. On the other hand, in our models, tidally
enhanced stellar wind in binary evolution naturally provides a physical mechanism for
mass loss on the RGB stage in late hot flash scenario, but we did not
consider the effects of stellar rotation on our final results, and it will be
studied in the near future.

\subsection{The role of binaries on formation of BHk stars in GCs}
As we can see from this study, the orbital periods needed to
produce BHk stars due to the tidally enhanced stellar wind
in binaries are relative long (e.g., from about 900 to 4600 d depends on the
models list in Table 1) when comparing with close binaries, thus  these
binaries would not be found easily in GCs from observation due to their
long orbital periods.  Moreover, due to the dense environment in GCs,
soft binaries (i.e., binaries with longer orbital periods) woud be
destroyed easily during the dynamical evolution of GCs (Hut et al. 1992;
Hurley, Aarseth \& Shara 2007; Heber 2016).  The boundary between hard and soft
binaries depends on the mean kinetic energy of the cluster stars (Hut et al. 1992).
A useful method was adopted by Hurley, Aarseth \& Shara (2007) to estimate
the boundary between hard and soft binaries in terms of binary separation,
which is given by twice the cluster half-mass radius divided by $N$.
In this method, $N$ is the total number of cluster members, including single and binary stars.
We used this method to estimate the boundary  of GC $\omega$ Cen as well, which
was compared with our theoretical results in this study.
From the catalogue compiled by Harris (1996, version 2003), the half-mass radius
of $\omega$ Cen is about
4.18 arcmin; considering that its distance from the sun is about 5.3 kpc, we could obtain the half-mass
radius for this GC is to be 6.44 pc. By assuming $N=10^{6}$ in this
GC, we estimate the boundary between
hard and soft binaries in terms of separation is about 2.66 au
(Here au is the abbreviation of 'astronomical unit'). This separation corresponds to
an orbital period of about 1400 d for a binary where the masses of primary and secondary
are 0.83 and 0.52$M_{\odot}$, respectively.  If this is the case,
at least part of binaries in our model which  produce BHk stars could
be classified as hard binaries and would survive during the evolution of GC.
However, we also estimated a hard/soft boundary of about 1.46 au in GC NGC 2808 using the
same method and it corresponds to an orbital period of about 500 d for
a binary system with the mass of two components are 0.83 and 0.52$M_{\odot}$, respectively.
This boundary is much shorter than the orbital periods in our models needed
to produce BHk stars.  Note that, the orbital periods needed to produce
BHk stars in our models become shorter with a smaller tidal enhancement efficiency
(i.e., $B\rm_{w}$) or a larger ratio of primary to secondary (see figs 10 and 11 in this study).
Therefore, this discrepancy could be resolved by using a smaller $B\rm_{w}$ and a larger
ratio of primary to secondary in our models.

Castellani et al. (2006) found a peculiar group of stars, which are called HBp stars in their study,
near the HB of GC NGC 2808 using the
WFPC2 camera on board the Hubble Space Telescope.
These peculiar stars locate in the region of blue straggler
stars in optical CMD of NGC 2808, but very close to the
hot HB stars in ultraviolet CMD (see fig 2 in their study).
They proposed that this phenomena is  possibly caused by
the composite nature of the spectrum which probably results from
the photometric blends of  physical binarity (e.g, the faint companion
stars have not been detected;  see Allard et al. 1994).  By assuming
the blended star (the faint companion star in a binary) as an MS
star,  Castellani et al. (2006) found that a hot HB star could be
shifted to the HBp region in the CMD due to  a blend with a redder companion, and
the faintest and hottest HB stars presented the maximum shift. Their results
indicated that at least some of the blue and hottest HB stars in NGC 2808 are binaries.

Due to the long orbital periods, the binaries consisting of BHk stars predicted by
our models are
difficult to be found directly in GCs from the observation,
but there may be some indirect evidences to test our
scenario. Following the
results we presented here,  if the initial orbital periods of the binaries
shorter than the minimum values needed to produce BHk stars in GCs (e.g.,
the orbital periods labelled by $P_{3}$ in Table 1), the primary stars in
these binaries would fail to ignite helium burning in their cores and die
as helium WDs. On the other hand, if the initial orbital periods are short enough,
the primary star in a binary would fill its Roche lobe radius and transfer
mass to the secondary through Roche lobe overflow (RLOF). This kind
of mechanism also would form hot EHB stars (see Han et al. 2002, 2003).
However, when tidally enhanced stellar wind is considered into
binary evolution, the initial orbital period which is required for the occurrence of RLOF in binaries
would become much shorter. In the study of Tout \& Eggleton (1988),
when tidally enhanced stellar wind is not considered in binary evolution, the
initial orbital periods required for RLOF in binaries are nearly shorter than  1500 d for a primary
star with a mass of 2 $M_{\odot}$, depending on the mass ratio adopted. However, the required initial orbital periods for
RLOF decrease to
less than 30 d if tidally enhanced stellar wind is considered (see figs 1 and 2 in their study),
and the final orbital periods (e.g., at the point when RLOF begins in a binary)
for these binaries are nearly less than 100 d (see fig 3 in Tout \& Eggleton 1988).
For a less massive
primary star (i.e., 1.3 $M_{\odot}$), these initial orbital periods required for
RLOF become even more shorter, e.g., less than 10 d (see fig 4 in Tout \& Eggleton 1988).
Therefore, based on these results and the results we presented in this
study, helium WD binaries would be formed through tidally enhanced stellar wind
in our models if the initial orbital periods are
in the range of  several tens to hundreds of days. Thus these kind of binaries
present much shorter orbital periods than the binaries consisting of
BHk stars predicted by our models, and they would be detected more easily from the observations.
If these kinds of binaries are found in GCs, it is an indirect evidence to support our
scenario presented here.

Actually, Kaluzny et al. (2007, 2013, 2014, 2015) have done
a series of study on detached eclipsing binaries (DEBs) in GCs.
They aimed to obtain the masses, orbital periods, radius and luminosity
of these binary components with higher precision, finally to determine
the ages and distances of the GCs which could be
used to test the stellar evolution models (Kaluzy et al. 2005).
Among the DEBs listed in these literatures, most of the binaries present
short orbital periods (e.g., several days to more than 10 d), and the
two components of these binaries are usually MS stars with nearly equal masses.
However, Kaluzny et al. (2013) found a DEB named V69 in GC M4, which
presents an orbital period of 48.19 d. The two components are MS stars near the
turn-off point in the CMD of M4, with a mass of $0.7665\pm0.0053 M_{\odot}$ for the primary and
a mass of $0.7278\pm0.0048 M_{\odot}$ for the secondary. This
binary is a possible candidate system to form the
helium WD binaries through tidally enhanced stellar wind.
Following our scenario, the primary star in this binary system
would be probably lose huge envelope mass at the RGB stage through
tidally enhanced stellar wind rather than through RLOF,
and become a helium WD in a binary system with a orbital period of several ten days to more than 100 d.

\section{Conclusions}
 As a further study for Paper I, by considering tidally enhanced
stellar wind into binary evolution, we studied in detail
the effects of metallicity, mass ratio, tidal enhancement
efficiency and helium abundance  on the formation of BHk stars .
Over 20 sets of binary models combined with different values of parameters mentioned above
are adopted to study their effects on the formation of BHk stars.
For each set of models, the space range of  initial  orbital periods which are
needed to produce BHk stars for binaries  was presented. We also
showed the evolution tracks of primary stars from ZAMS to WD including the
stage of late hot helium flash,   as well
as the evolution parameters on the ZAHB for some primary stars.

Our results are compared with the observation in $\it{T}\rm_{eff}$-$\rm{log}\it{g}$
plane. Even though the input parameters of these binary models are
different from each other, all  binaries in  these models could produce
BHk stars within different range of initial orbital periods. Most of the BHk stars
in NGC 2808 and $\omega$ Cen locate well  in the region predicted by
our models, especially when C/N-enhanced model atmospheres are considered in
obtaining the stellar atmosphere parameters of BHk stars.
The effects of metallicity, mass ratio, tidal enhancement efficiency and
helium abundance on the space range of initial orbital periods needed
to produce BHk stars are discussed in detail.
We found that mass ratio of primary to secondary and tidal enhancement efficiency
have little effects on the formation of BHk stars in our model, while metallicity and
helium abundance seem to play important roles in formation of
BHk stars in binaries. Especially,  with the helium abundance increasing
from $Y$=0.24-0.40, the range of initial orbital period to produce
BHk stars becomes wider obviously for all the models. Assuming a flat
initial orbital periods distribution for binaries in GCs, one would expect
a much easier production of BHk stars if these stars have higher initial
helium abundance. Our results presented here indicate that tidally
enhanced stellar wind in binary evolution is a possible formation channel for
BHk stars in GCs, but further studies and more evidences are needed to investigate the roles
of binaries on this problem before it comes to a conclusive result.

\section*{Acknowledgements}
We thank the anonymous referee for valuable comments
and suggestions that helped us to improve the paper.
This work is supported by the National Natural Science Foundation
of China (Grant Nos, 11503016, 11390371, 11233004, 11573061),
the Youth Fund project of Hunan Provincial Education Department
(Grant No. 15B214). ZH is partly supported by the
Natural Science Foundation of China (Grant No. 11390374, 11521303), the
Science and Technology Innovation Talent Programme of the Yunnan
Province (Grant No. 2013HA005) and the Chinese Academy
of Sciences (Grant No. XDB09010202, KJZD-EW-M06-01).

\label{lastpage}

\end{document}